\documentstyle[prb,aps,psfig,floats]{revtex}
\begin{document}
\twocolumn
\draft
\wideabs{
\title
{Superconductivity in the Re-B system}

\author{G.~K.~Strukova, V.~F.~Degtyareva, D.~V.~Shovkun, V.~N.~Zverev, V.~M.~Kiiko,
A.~M.~Ionov, and A.~N.~Chaika}
\address{Institute of Solid State Physics, 142432, Chernogolovka,
Moscow district, Russia}

\date\today
\maketitle
%-----------------------------------------------
\begin{abstract}
Superconductivity was found for a rhenium boride Re$_3$B at
$T_c$=4.7~K and for ReB$_2$ with $T_c$ in the range from 4.5 to
6.3~K depending on the boron concentrations. Both compounds have
the structure different from that of the simple layered diborides,
in particular from MgB$_2$.
\end{abstract}
%-----------------------------------------------
}
\thispagestyle{empty}
\newpage
%--------------------------------------------------------
Since the discovery of superconductivity in magnesium diboride
many works have been focused on investigation of superconductivity
in diborides with the structure of AlB$_2$ type. This structure is
a characteristic of some superconductive diborides of transition
metals \cite{Kaczorowski}.

Our attention has been attracted by rhenium borides with the
structure different from that of the simple layered diborides, in
particular from MgB$_2$. In the Re-B system following compounds
are known Re$_3$B, Re$_7$B$_3$, ReB$_2$ and ReB$_3$ \cite{binary}
with the structures different from the AlB$_2$ type.
Superconductivity at $T_c$=2.8~K was found for a rhenium boride
referred to Re$_2$B (with undetermined structure) \cite{Hulm}.
Therefore it is interest to check the superconduting properties of
the Re-B compounds and to correlate them with the crystal
structure.
%fig1.----------------------------------------
\begin{figure}[h]
\centerline{\psfig{figure=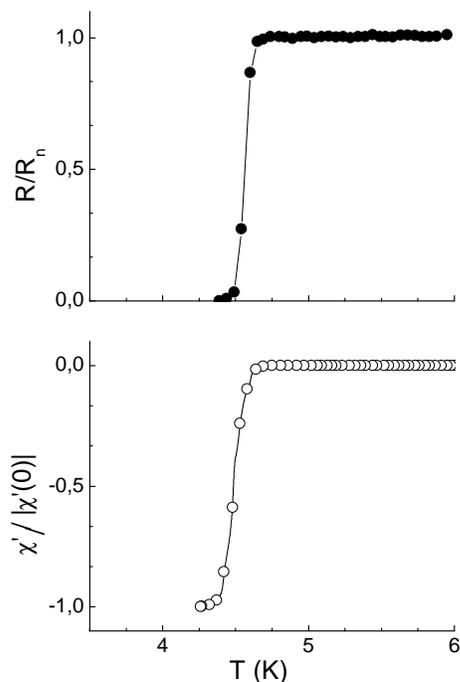,height=9cm,width=6cm,clip=,angle=0.}}
\caption{Temperature dependence of the resistance normalized by
the normal state resistance near $T_c$ (upper curve) and the
normalized ac-susceptibility (lower curve) of Re$_3$B sample.}
\label{f1}
\end{figure}

The samples of rhenium diborides were synthesized by heating of
rhenium with amorphous boron at temperatures from 950 to
1000$^\circ$C in an inert atmosphere. An increase in the synthesis
temperature causes a decomposition of rhenium diboride onto
metallic rhenium and Re$_3$B phase.

The measurements of the ac-susceptibility and resistance
temperature dependence shown, that both compounds (ReB$_2$ and
Re$_3$B) are superconductors at temperatures slightly above liquid
helium. The temperature of transition into superconductive state
for ReB$_2$ is in the range from 4.5 to 6.3~K depending on the
boron concentrations. Measurements of the sample contents by Auger
electron spectra shown boron content from 2 to 1.8.

The Re$_3$B compound displays the temperature of superconductive
transition about 4.7~K. Fig.1 illustrates the temperature
dependence of the ac-susceptibility and resistivity for this
compound. Diffraction pattern from the powdered sample corresponds
to the known Re$_3$B compound with the orthorhombic cell, space
group {\it Cmcm}, 16 atoms in the cell. Measured lattice
parameters are a=2.8905, b=9.3039, c=7.2641~\AA ($\pm$~0.0003) in
agreement with the literature data \cite{Villars}.

\end{document}